\documentclass[11pt]{article}

\usepackage[preprint]{acl}

\usepackage{times}
\usepackage{latexsym}
\usepackage{booktabs}
\usepackage{amssymb}

\usepackage[T1]{fontenc}

\usepackage[utf8]{inputenc}

\usepackage{microtype}

\usepackage{inconsolata}

\usepackage{graphicx}

\usepackage{amsmath}
\usepackage{booktabs}
\usepackage{multirow}

\usepackage{pifont}
\usepackage{arydshln}
\usepackage{marvosym}
\usepackage{fontawesome5} 
\usepackage{hyperref}

\title{HyTRec: A Hybrid Temporal-Aware Attention Architecture for Long Behavior Sequential Recommendation}

\author{
  \textbf{Lei Xin\textsuperscript{\rm 1,2}},
 \textbf{Yuhao Zheng\textsuperscript{\rm 3}},
 \textbf{Ke Cheng\textsuperscript{\rm 4}},
 \textbf{Changjiang Jiang\textsuperscript{\rm 2}},
 \textbf{Zifan Zhang\textsuperscript{\rm 2}},
 \textbf{Fanhu Zeng \Letter} \\
 {\normalsize \textsuperscript{1}Shanghai Dewu Information Group} \
 {\normalsize \textsuperscript{2}Wuhan University} \
 {\normalsize \textsuperscript{3}USTC} \ 
 {\normalsize \textsuperscript{4}Beihang University}\\
 {\normalsize i\_xinlei@dewu.com, \ yuhaozheng@mail.ustc.edu.cn, \ kecheng@tencent.com,} \\ {\normalsize jiangcj@whu.edu.cn, \ zifan623@gmail.com, \ challengezengfh@gmail.com}
}

\begin{document}
\maketitle
\footnotetext[1]{\Letter \ Corresponding Author.}

\begin{abstract}
Modeling long sequences of user behaviors has emerged as a critical frontier in generative recommendation. 
However, existing solutions face a dilemma: linear attention mechanisms achieve efficiency at the cost of retrieval precision due to limited state capacity, while softmax attention suffers from prohibitive computational overhead.
To address this challenge, we propose \textbf{HyTRec}, a model featuring a \textbf{Hybrid Attention} architecture that explicitly decouples long-term stable preferences from short-term intent spikes. By assigning massive historical sequences to a linear attention branch and reserving a specialized softmax attention branch for recent interactions, our approach restores precise retrieval capabilities within industrial-scale contexts involving ten thousand interactions. To mitigate the lag in capturing rapid interest drifts within the linear layers, we furthermore design \textbf{Temporal-Aware Delta Network (TADN)} to dynamically upweight fresh behavioral signals while effectively suppressing historical noise. Empirical results on industrial-scale datasets confirm the superiority that our model maintains linear inference speed and outperforms strong baselines, notably delivering over 8\% improvement in Hit Rate for users with ultra-long sequences with great efficiency.
\end{abstract}

\section{Introduction}
The rapid accumulation of interaction data on online platforms has driven a fundamental transition in recommender systems from traditional collaborative filtering to generative paradigms rooted in ultra-long sequences \cite{zhai2024actions,lin2025survey,deng2025onerec,du2025mom,tabdsr}. Central to this shift is the utilization of long behavior sequences, which serve as a vital resource for decoding dynamic preferences and latent intentions \cite{kang2018self,gu2024mamba}. Such extensive sequences provide rich interaction signals that offer high-quality feedback for the next item prediction task \cite{sun2024learning,fakehr1}. By revealing the long-term trajectory of user interests rather than just short-term session data, modeling long behavior sequences proves essential for capturing complex decision-making paths.

To model these sequences, methodologies have evolved significantly. Early approaches utilized session-based patterns \cite{hidasi2016session} or self-attentive models \cite{kang2018self,sun2019bert}, establishing the foundation for sequential modeling.
More recently, the field has witnessed the rise of generative frameworks, represented by P5 \cite{geng2022recommendation} and TALLRec \cite{bao2023tallrec}, which demonstrate robust generalization capabilities.
Furthermore, emerging simulators \cite{wang2025simuser} have highlighted the value of modeling long sequence data.

\begin{figure}[t]
  \centering
  \includegraphics[width=\linewidth]{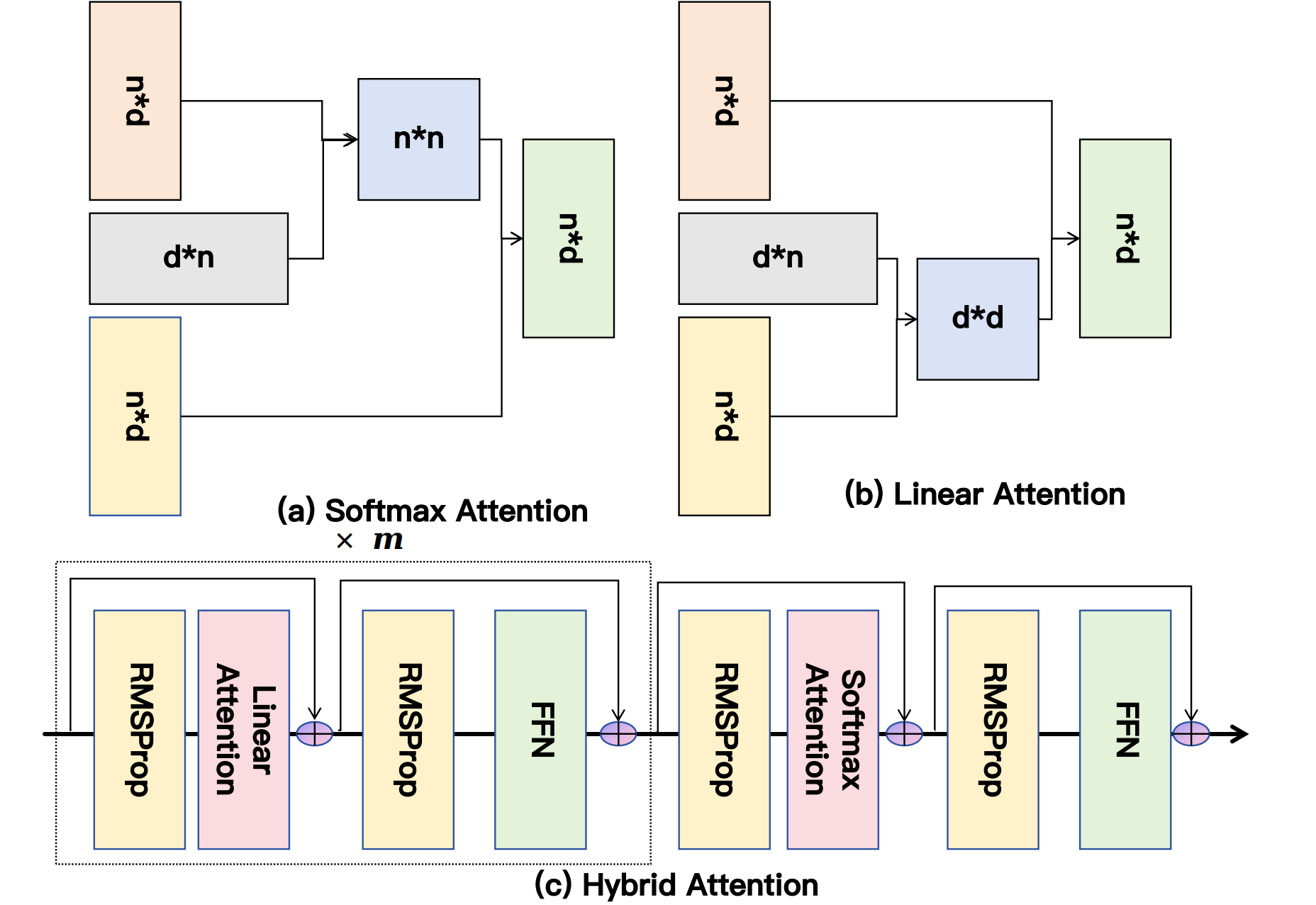}
  \caption{The Evolution of Attention Mechanisms.}
  \label{fig:Attention}
\end{figure}

Nevertheless, existing models encounter two formidable obstacles when scaling to long behavior sequences:
\textit{\textbf{(i)}} The inherent trade-off between efficiency and expressiveness remains unresolved, as traditional softmax attention suffers from quadratic complexity while linear variants often compromise retrieval precision, resulting in semantic ambiguity and limited injectivity when capturing fine-grained dependencies \cite{qin2025bridging}, as shown in Figure \ref{fig:Attention}.
\textit{\textbf{(ii)}} Current architectures struggle to adapt to interest drifts \cite{zhou2018deep}. Linear models \cite{gu2023mamba,yang2024parallelizing} often fail to catch up with rapid intent changes because they compress all information into a fixed state. Consequently, they cannot easily distinguish immediate, high-value signals from the vast amount of historical noise, leading to a lag in capturing what the user truly wants at the moment \cite{shao2025motir}.

To address these limitations, we propose \textbf{HyTRec}, a generative framework featuring a \textbf{Hy}brid \textbf{T}emporal-Aware \textbf{Rec}ommendation architecture tailored for efficient long behavior sequence modeling. 
\textit{\textbf{(i)}} To reconcile the trade-off between inference speed and retrieval precision, we design a \textbf{Hybrid Attention} architecture specifically for modeling long behavior sequences. By strategically integrating a small proportion of softmax attention layers into a predominantly linear attention backbone, we maintain near-linear complexity comparable to purely linear models, while effectively restoring the high-fidelity retrieval capabilities that are typically compromised in linear approximations.
\textit{\textbf{(ii)}} To further mitigate the lag in capturing rapid interest drifts within the linear layers, we incorporate a \textbf{Temporal-Aware Delta Network (TADN)}. This module utilizes an exponential gating mechanism to dynamically upweight fresh behavioral signals, effectively suppressing historical noise and ensuring the model remains highly sensitive to immediate user intents.
Extensive evaluations on diverse benchmarks validate the effectiveness of HyTRec, where it consistently outperforms strong baselines by an average of 5.8\% in NDCG.
Our contributions are as follows:
\begin{itemize}
    \item \textbf{Novel Hybrid Attention.} 
    We propose HyTRec, a hybrid attention framework for generative recommendation that synergizes linear attention for history with softmax attention for recent interactions, achieving linear complexity while preserving the semantic integrity of long-term preferences.

    \item \textbf{Dynamic Intent Modeling.} 
    We introduce the Temporal-Aware Delta Networks (TADN), which leverage a temporal decay factor to meticulously track rapid interest shifts, ensuring transient user intents are accurately prioritized over historical noise.

    \item \textbf{Empirical Performance.}
    Extensive experiments on real-world e-commerce datasets demonstrate that HyTRec outperforms strong baselines, delivering over 8\% improvement in Hit Rate for users with extensive interaction histories while maintaining linear inference speed.
\end{itemize}

\section{Related Work}

\subsection{Sequential and Generative Recommendation} 
Sequential recommendation aims to predict subsequent user actions by characterizing dynamic dependencies within historical interactions. Early methodologies primarily utilized Markov Chains for short-term transitions or Recurrent Neural Networks to capture broader sequential regularities. With the rise of Transformer architectures, models such as SASRec \cite{kang2018self} leveraged self-attention for global dependency modeling. In industrial practice, managing ultra-long behavior sequences led to the development of SIM \cite{pi2020search}, which employs a two-stage search-based strategy, and ETA \cite{chen2021end}, which utilizes Locality Sensitive Hashing for end-to-end processing.

Recently, the paradigm has shifted toward generative recommendation inspired by Large Language Models \cite{zhai2024actions, lin2025survey}. In this context, P5 \cite{geng2022recommendation} unified multiple recommendation tasks, while TablePilot \cite{liu2025tablepilot} demonstrated the feasibility of aligning generative models with human-preferred data analysis patterns. To address evaluation constraints, SimUSER \cite{wang2025simuser} explored using LLMs to simulate user behavior. For cold-start scenarios, reinforcement learning-driven adversarial query generation has been employed to enhance relevance \cite{shukla2025reinforcement}. Furthermore, to capture diverse evolving user intentions, MIND \cite{li2019multi} and ComiRec \cite{cen2020controllable} introduced multi-interest extraction mechanisms. Despite these advancements, the practical deployment of generative recommendation is severely constrained by inference latency. The substantial computational overhead required to process lifelong sequences with Large Language Models often exceeds the strict response-time limits of real-time industrial systems, necessitating more efficient architectural solutions.

\subsection{Efficient Long-Sequence Modeling and Hybrid Architectures}
The evolution of long-sequence modeling provides a strategic roadmap for processing lifelong behavioral data. Although the original self-attention mechanism \cite{vaswani2017attention} offers high expressiveness, its computational complexity limits practical scalability. Consequently, Longformer \cite{beltagy2020longformer} introduced sparse attention patterns, and linear variants like Performers \cite{choromanski2020rethinking} utilized kernel-based approximations. Subsequently, State Space Models such as S4 \cite{gu2021efficiently} and Mamba \cite{gu2023mamba}, along with linear variants like DeltaNet \cite{yang2024parallelizing}, achieved strict linear complexity $O(n)$.

However, pure linear models frequently suffer from semantic confusion and struggle to maintain the injectivity of their hidden state updates \cite{qin2025bridging}. Moreover, standard embedding techniques in ultra-long sequence scenarios can lead to information bottlenecks, necessitating the use of decoupled embeddings to preserve model capacity \cite{ren2025long}. To strike a balance between efficiency and precision, the industry is exploring hybrid architectures where frameworks like Jamba \cite{lieber2024jamba} interleave state-space layers with self-attention layers. HyTRec advances this philosophy by introducing the Temporal-Aware DeltaNet. Techniques such as ALiBi \cite{press2021train} utilize static linear biases to weight local context effectively, whereas MotiR \cite{shao2025motir} emphasizes the retrieval of underlying user motivations to filter out superficial noise. Building upon these distinct approaches, our model explicitly incorporates time-decay factors via an exponential gating mechanism to dynamically prioritize recent high-intent behaviors. Unlike standard Gated Linear Attention \cite{yang2024gated}, this design enables the model to discard irrelevant distant history while focusing on recent high-intent signals\textbf{, thereby effectively resolving the semantic dilution problem common in long-sequence modeling.}

\section{Preliminaries}
In this work, we focus on the task of next-item prediction within a long sequential recommendation scenario. The core objective is to accurately predict the ID of the next item that a user will purchase based on their historical interaction sequence.

Let $\mathcal{U}$ denote the set of users, and $\mathcal{I}$ denote the set of items. For an arbitrary user $u \in \mathcal{U}$, the long interaction sequence is represented as $S_u = [x_1, x_2, \dots, x_n]$, where $x_t \in \mathcal{I}$ denotes the item interacted with at time step $t$, and $n$ represents the length of the user's interaction sequence. 

Formally, the goal of the recommender system is to estimate the probability distribution of the next item $x_{n+1}$ given the historical sequence $S_u$, formulated as maximizing $P(x_{n+1} \mid S_u)$.

\section{HyTRec}

\begin{figure*}[t]
  \centering
  \includegraphics[width=\textwidth]{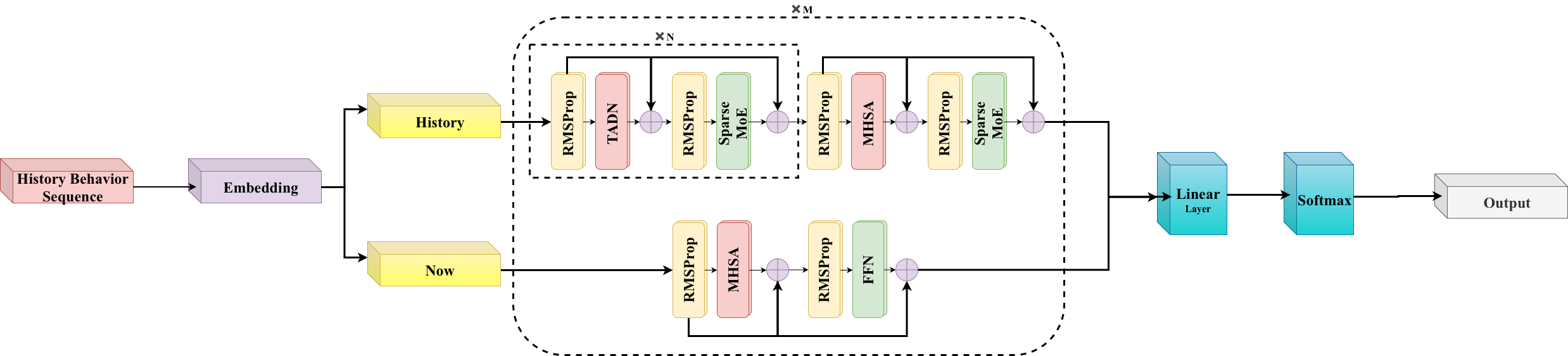}
  \caption{The Framework of HyTRec.}
  \label{fig:framework}
\end{figure*}

\subsection{Framework Overview}
We propose HyTRec, a generative framework designed for long sequence users behavior modeling. As illustrated in Figure \ref{fig:framework}, the framework explicitly decouples the processing of massive historical patterns from immediate intent spikes from sequence and architecture design, respectively.

\noindent \textbf{Sequence Decomposition Strategy.}
To achieve this decoupled modeling, we first decompose the long sequence $S_u$ into two disjoint subsequences:
\begin{itemize}
    \item \textbf{Short-term behavior sequence}.
    The length of this sequence is fixed at $K$, denoted as $S_u^{short} = [x_{n-K+1}, \dots, x_n]$. This subsequence focuses on the user's recent behaviors to capture short-term sudden consumption intents and interest drifts.
    \item \textbf{Long-term historical behavior sequence}.
    This sequence consists of the historical interaction excluding short-term part, denoted as $S_u^{long} = [x_1, \dots, x_{n-K}]$, with a length of $n-K$. This subsequence covers long-term behavioral patterns to capture stable and inherent consumption preferences.
\end{itemize}

\noindent \textbf{Dual-Branch Data Flow.}
Building upon this stratification, HyTRec processes these signals through two parallel branches before fusion:
\begin{itemize}
\item The \textbf{short-term branch} processes $S_u^{short}$ using standard multi-head self-attention (MHSA) to ensure maximum precision for recent behaviors.
\item The \textbf{long-term branch} processes $S_u^{long}$ using our proposed hybrid attention architecture. This branch serves as the computational backbone, compressing the extensive history into a compact representation while preserving fine-grained dependencies. 
\end{itemize}
The outputs of both branches are subsequently fused to generate the final prediction.

\subsection{Hybrid Attention Architecture}
Building upon above decomposition, a critical innovation of HyTRec lies in the long-term branch, where we introduce a \textit{hybrid attention mechanism} to break the $O(n^2)$ complexity bottleneck while maintaining global context awareness.
The core design philosophy of hybrid attention is to decouple the modeling of extensive history from recent interactions to \textit{balance efficiency and retrieval precision}.
Unlike traditional methods that rely solely on either efficient linear attention (suffering from recall loss) or heavy softmax attention (suffering from computational overhead), we design a hybrid layer stack. Specifically, the long-term branch is composed of $N$ encoder layers, predominantly utilizing our novel Temporal-Aware Delta Network (TADN) as the base unit for linear complexity, while strategically interleaving a small proportion of standard attention layers (\textit{e.g.}, at a ratio of 7:1). This strategy significantly improves computational efficiency (compared to a pure standard attention layer) while preserving model expressiveness and long-range modeling capabilities.

\subsection{TADN: Temporal-Aware Delta Networks}
To precisely capture the temporal dynamics of user purchasing behaviors while maintaining long-term preferences, TADN introduces a Temporal-Aware Gating Mechanism. By quantifying the correlation between historical behaviors and the target purchase via temporal decay, the model achieves an effective balance between short-term deviations and long-term inherent preferences.

\paragraph{Temporal Decay Factor.}
We define the temporal decay factor $\tau_t$ to measure the relevance of a past interaction to the current decision:
\begin{equation}
    \tau_t = \exp\left(-\frac{t_{\text{current}} - t_{\text{behavior}}^t}{T}\right),
\end{equation}
where $t_{\text{behavior}}^t$ denotes the timestamp of the historical behavior, $t_{\text{current}}$ is the timestamp of the next purchase action, and $T$ represents the decay period. Here, $\tau_t \in (0, 1]$ directly characterizes the correlation strength.

\paragraph{Temporal-Aware Gating Generation.}
Fusing the temporal decay factor with feature similarity, we generate dynamic gating weights $g_t$ to highlight the impact of recent behaviors:
\begin{equation}
\begin{aligned}
    g_t = \alpha \cdot [ \sigma(\mathbf{W}_g &\cdot \text{Concat}(\mathbf{h}_t, \Delta \mathbf{h}_t) \\
    + &\mathbf{b}) \odot \tau_t ] + (1 - \alpha) \cdot g_{\text{static}},
\end{aligned}
\label{eqn:weight}
\end{equation}
where $\Delta \mathbf{h}_t = \mathbf{h}_t - \bar{\mathbf{h}}$ represents short-term preference features, $g_{\text{static}} = \text{softmax}(\mathbf{h}_t^\top \bar{\mathbf{h}} / \sqrt{d})$ serves as the static gate for long-term preferences, and $\alpha$ is the balancing coefficient.

\paragraph{Information Fusion Mechanism.}
Based on the generated gate, we dynamically amplify the contribution of behaviors highly correlated with the recent purchase period while preserving long-term preferences:
\begin{equation}
    \tilde{\mathbf{h}}_t = g_t \odot \Delta \mathbf{h}_t + (\mathbf{1} - g_t) \odot \mathbf{h}_t.
\end{equation}
The output is the fused feature matrix $H = [\tilde{\mathbf{h}}_1, \dots, \tilde{\mathbf{h}}_L]^\top$. In this mechanism, behaviors closer to the purchase time obtain higher gating weights due to a larger $\tau_t$. Consequently, the short-term deviation feature $\Delta \mathbf{h}_t$ dominates the output, enabling a rapid response to immediate purchase intents (\textit{e.g.}, in flash sales scenarios). Conversely, for routine behaviors, long-term preferences are effectively preserved via the $g_{\text{static}}$ component.

\paragraph{Temporal-aware Gated Delta Rule}
Finally, we integrate the fused features into the Gated DeltaNet framework. By substituting the temporal-aware gate $g_t$ into the state update rule, we derive the TADN linear attention formulation.
Let $\mathbf{S}_t \in \mathbb{R}^{d \times d}$ be the hidden state matrix. The recurrence relation is defined as:
\begin{equation}
    \mathbf{S}_t = \mathbf{S}_{t-1} \left( \mathbf{I} - g_t \beta_t \mathbf{k}_t \mathbf{k}_t^\top \right) + \beta_t \mathbf{v}_t \mathbf{k}_t^\top,
\end{equation}
where $\beta_t \in (0, 1)$ is the writing strength from the standard delta rule, and $\mathbf{k}_t, \mathbf{v}_t$ are the key and value vectors derived from the input features.

By expanding this recurrence, we derive the formulation of linear attention for TADN. The state $\mathbf{S}_t$ can be expressed as a summation of historical updates weighted by a cumulative decay path:
\begin{equation}
    \mathbf{S}_t = \sum_{i=1}^t \left( \prod_{j=i+1}^t (\mathbf{I} - g_j \beta_j \mathbf{k}_j \mathbf{k}_j^\top) \right) \beta_i \mathbf{v}_i \mathbf{k}_i^\top.
\end{equation}
Consequently, the final output $\mathbf{o}_t = \mathbf{S}_t \mathbf{q}_t$ (which corresponds to the fused feature $\tilde{\mathbf{h}}_t$ in our context) can be formulated as a linear attention operation with a temporal-aware decay mask:
\begin{align}
\mathbf{o}_t = \sum_{i=1}^t \beta_i \left( \mathbf{v}_i \mathbf{k}_i^\top \right) \mathbf{q}_t \cdot \mathcal{D}(t, i), \\
\mathcal{D}(t, i) = \prod_{j=i+1}^t \left(\mathbf{I} - g_j \beta_j \mathbf{k}_j \mathbf{k}_j^\top\right),
\end{align}
where $\mathcal{D}(t, i)$ represents the composite decay mask from time $i$ to $t$. Unlike standard Gated DeltaNet where decay is purely semantic, in TADN, the term $g_j$ explicitly contains the temporal factor $\tau_j$ (from Eq.~\ref{eqn:weight}). This ensures that the attention mechanism mathematically prioritizes recent interactions (where $\tau \approx 1$) while preserving long-term preferences through the static component of $g_j$.
\begin{table}[htbp]
  \centering
  \caption{Cross-domain transfer experiments based on the 2022 Huawei Advertising Challenge Dataset. The table presents the performance on Recall@10, GAUC and AUC metrics.}
  \resizebox{0.9\linewidth}{!}{\begin{tabular}{lccc}
    \toprule
    \textbf{Model} & \textbf{Recall@10} & \textbf{GAUC} & \textbf{AUC} \\
    \midrule
    SASRec & 0.0235 & 0.4224 & 0.4533 \\
    \textbf{Ours} & \textbf{0.0317} & \textbf{0.8758} & \textbf{0.9327} \\
    \bottomrule
  \end{tabular}}
  \label{tab:cross-domain}
\end{table}

\section{Experiments}
We structure our experimental analysis to investigate four core research questions:

\noindent \textbf{RQ1:} How does HyTRec perform against state-of-the-art baselines in modeling long user behaviors?

\noindent \textbf{RQ2:} Can HyTRec effectively scale to ultra-long sequences while maintaining competitive training and inference efficiency?

\noindent \textbf{RQ3:} How does each key component of HyTRec contribute to the overall performance?

\noindent \textbf{RQ4:} How does the hybrid attention ratio affect retrieval precision and system latency?

\noindent \textbf{RQ5:} How robust is HyTRec in handling specific challenging scenarios, such as the cold-start phase for new users?

\subsection{Experimental Setup}
\paragraph{Datasets.}
We evaluate on four widely-used recommendation benchmarks: Amazon Beauty, Amazon Movies \& TV, and Amazon Electronics. Recommendation systems are selected as a stress test for High-Rank Sparsity because user interests are inherently multimodal and diverse, which presents a sharp contrast to low-rank attention mechanisms commonly found in natural language processing.

To mitigate the impact of new-user cold starts and inactive-user silence on long behavior sequence modeling, we filtered the data based on interaction frequency and product recurrence counts.

\paragraph{Setup and Evaluation.}
We employ users' historical behavior sequences to predict the next item they will purchase. Importantly, we compare different methods under roughly matched computational budgets: the per-sample sequence FLOPs and per-step runtime are aligned across methods. For the quadratic-cost encoders (Transformer and HSTU), we reduce their depth and width so that their overall computation is comparable to Stacked
Target-to-History Cross Attention (STCA), ensuring a fair comparison. We compare single-layer target attention, DIN, Transformer, and HSTU against our HyTRec, and adopt Request Level Batching (RLB) with sparse training and dense inference.

\paragraph{Baselines.}
We compare our proposed HyTRec method with behavior sequence models and long-text models as baselines. For the former, we compare GRU4Rec, SASRec, DIN, and HSTU. For the latter, we compare Transformer, GLA, and Qwen-next (2 blocks). We conduct extensive experiments verify the superiority of our method.

\paragraph{Metrics.}
We adopt H@500, NDCG@500, and AUC as metrics to assess the model’s performance on product recommendation, while latency is used to evaluate its response speed. All experiments are carried out on V100 GPUs.
\begin{table*}[t]
\centering
\caption{We compare with various methods on three public Amazon datasets in terms of H@500, NDCG@500 and AUC. The best and second-best methods are marked in bold and underlined, respectively.}
\setlength{\tabcolsep}{3.5pt}
\resizebox{\linewidth}{!}{
\begin{tabular}{lcccccccccc}
\toprule
\multirow{2}{*}{\textbf{Model}} & \multicolumn{3}{c}{\textbf{Beauty}} & \multicolumn{3}{c}{\textbf{Electronics}} & \multicolumn{3}{c}{\textbf{Movies\&TV}} \\
\cmidrule(lr){2-4} \cmidrule(lr){5-7} \cmidrule(lr){8-10}
                      & \textbf{H@500}  & \textbf{NDCG@500} & \textbf{AUC}      & \textbf{H@500}    & \textbf{NDCG@500} & \textbf{AUC}       & \textbf{H@500}    & \textbf{NDCG@500} & \textbf{AUC}       \\
\midrule
GRU4Rec               & 0.5263  & 0.1648   & 0.8397   & 0.2854   & 0.0752   & 0.8353    & 0.5380    & 0.1261   & 0.9307    \\
SASRec                & 0.5776  & 0.3280   & 0.8497   & 0.3003   & 0.0945   & 0.8625    & 0.6979   & 0.6226   & 0.9370     \\
HSTU                  & \underline{0.5838} & \textbf{0.4480} & \underline{0.8602} & 0.3156   & \textbf{0.1244} & 0.8550     & 0.7042   & \textbf{0.6278} & \underline{0.9372} \\
DIN                   & 0.5748  & \underline{0.4297} & 0.8556   & 0.2700   & 0.1063   & 0.8560    & \textbf{0.7155} & 0.6236   & 0.9263    \\
GLA                   & 0.5732  & 0.4157   & 0.8548   & 0.3176   & 0.1042   & 0.8657    & 0.7053   & 0.6153   & 0.9124    \\
Qwen-Next (2 block)   & 0.5807  & 0.4291   & 0.8598   & \textbf{0.3686} & 0.1065   & \textbf{0.8797} & 0.7060   & 0.5938   & \textbf{0.9419} \\
\textbf{HyTRec}       & \textbf{0.6643} & 0.3480   & \textbf{0.8655} & \underline{0.3272} & \underline{0.1192} & \underline{0.8760} & \underline{0.7070} & \underline{0.6268} & 0.9191    \\
\bottomrule
\end{tabular}
}
\label{tab:comparison_user_groups}
\end{table*}

\subsection{Performance Comparison (RQ1)}
The results on three public datasets are shown in Table \ref{tab:comparison_user_groups}. The common laws and core differences among the three types of models are clearly distinguishable: all three take capturing the correlation of user interests in long behavior sequences as their core goal, and the model performance is positively correlated with the context modeling ability and long-sequence adaptability. The core differences focus on the trade-off between efficiency and expressive ability. Specifically, transformer-based models (\textit{e.g.}, SASRec, HSTU) rely on full context modeling with quadratic complexity to become the performance upper bound, but they are difficult to adapt to ultra-long sequences; linear attention-based models (\textit{e.g.}, GRU4Rec) have high efficiency but limited expressive ability, resulting in significant performance gaps; hybrid attention-based models (including our HyTRec and Qwen-next) attempt to balance the two, showing better overall indicator performance, but there are differences in adaptability and comprehensive performance among different models.

As a result, as a representative model of hybrid attention models, our HyTRec shows significant advantages compared with its counterparts and the other two types of models: on the Beauty dataset, the H@500 (0.6643) of HyTRec is much higher than that of all three types of baselines, and its AUC (0.8655) ranks first, highlighting a stronger ability to capture user interests; on the Electronics dataset, its H@500 (0.3272) is second only to Qwen-next, a hybrid model of the same type, and its AUC (0.876) is better than all other baselines, adapting to long-sequence scenarios with scattered interests; on the Movies\&TV dataset, its H@500 (0.707) and NDCG@500 (0.6268) are close to the optimal level of transformer-based models, with stable and efficient performance. All the results strongly show the superiority of our method.

\begin{figure}[t]
  \centering
  \includegraphics[width=\columnwidth]{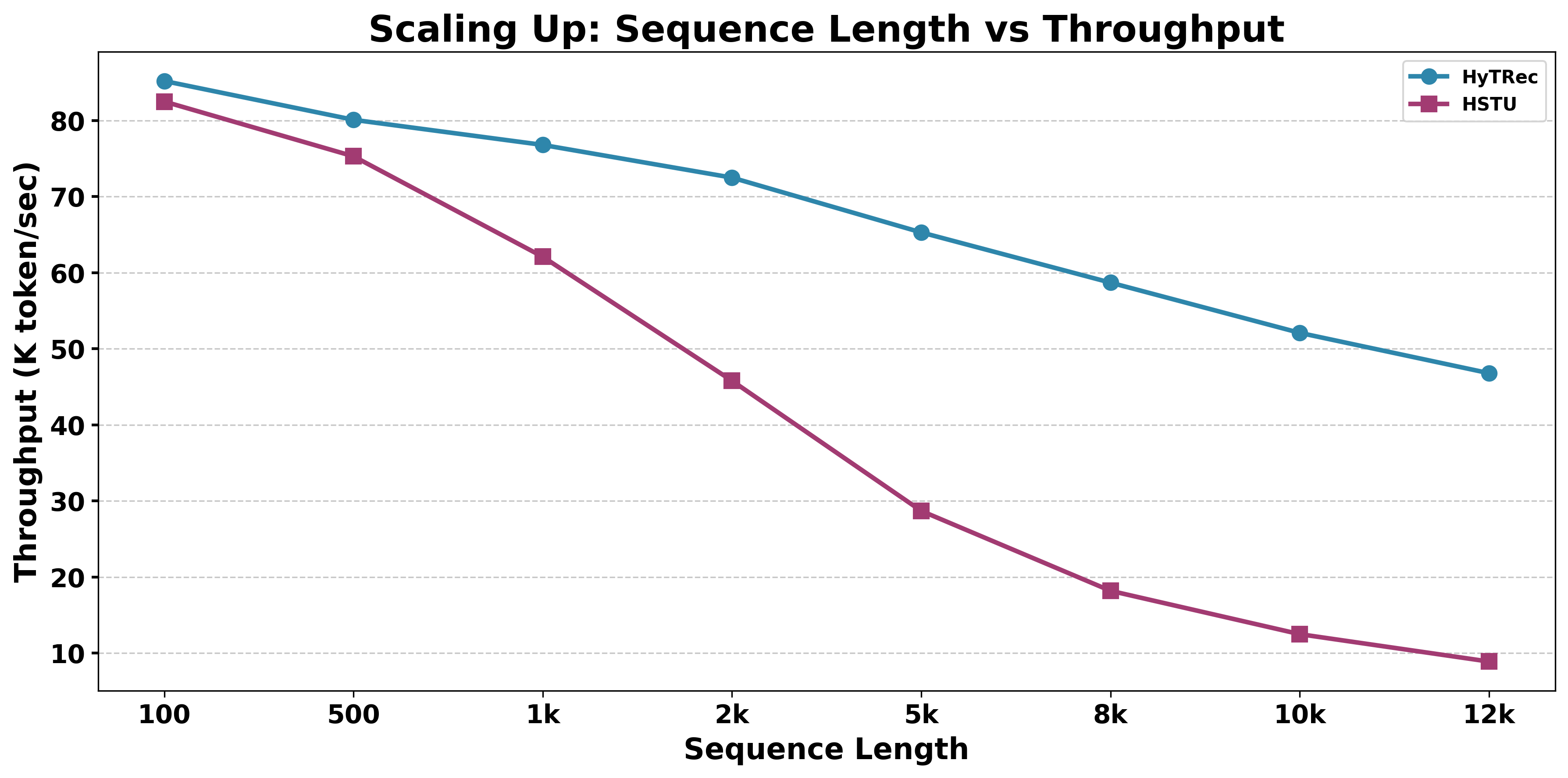}
  \caption{We compare the training throughput of models with the same parameter scale on a single V100 GPU under different behavior sequence lengths.}
  \label{fig:scaling_up}
\end{figure}

\begin{table*}[htbp]
  \centering
  \caption{Ablation experiments of the model on the Amazon Beauty dataset.}
  \begin{tabular}{lcccccc}
    \toprule
    \textbf{Model} & \textbf{TADN} & \textbf{Short-Term Attention} & \textbf{H@500} & \textbf{NDCG@500} & \textbf{AUC} \\
    \midrule
    \multirow{4}{*}{HyTRec} & $\times$ & $\times$ &0.6043  &0.3130  &0.8355  \\
                  & $\times$ & $\checkmark$ &0.6343  &0.3300  &0.8505  \\
                  & $\checkmark$ & $\times$ &0.6493  &0.3380  &0.8575  \\
                  & $\checkmark$ & $\checkmark$ &0.6643 &0.3480  &0.8655  \\
    \bottomrule
  \end{tabular}
  \label{tab:ablation_study}
\end{table*}

\subsection{Training Efficiency (RQ2)}
The performance curve shown in Figure \ref{fig:scaling_up} clearly demonstrates the core advantages of HyTRec in efficient long behavioral sequence modeling: across the entire range of sequence lengths from 100 to 12k, HyTRec maintains a steady downward trend in throughput. This characteristic stems from the linear attention mechanism adopted in the long-sequence branch of the model, where the linear computational complexity fundamentally circumvents the efficiency bottleneck caused by the increase in sequence length, enabling efficient processing of ultra-long behavioral sequences throughout the entire user lifecycle. In contrast, the HSTU model achieves a throughput close to that of HyTRec when the sequence length is $\leq$ 1k (85.2 K token/sec at the length of 100), but its throughput plummets drastically once the sequence length exceeds 1k: it drops to 28.7 K token/sec at the length of 5k and only remains at 8.9 K token/sec at the length of 12k, accounting for merely 19\% of HyTRec's throughput at the same sequence length. This is because the stacked sequence transduction units relied on by HSTU are essentially still constrained by quadratic complexity, making it unable to adapt to the modeling requirements of ultra-long behavioral sequences. 

As a typical length threshold for long behavioral sequences, HyTRec still sustains a high throughput of 65.3 K token/sec at 5k, while HSTU loses nearly 60\% of its processing efficiency. This fully verifies the practicability and scalability of HyTRec in industrial-level ultra-long behavioral sequence scenarios, and also proves the rationality of the design of the parallel dual-attention branch architecture in the integration of efficient long-sequence modeling and accurate short-term intent capture.

\subsection{Ablation Study (RQ3)}
To verify the effectiveness of each key component in our proposed efficient long behavioral sequence model, we conduct ablation experiments on the recommendation task, with HyTRec as the baseline model. The experimental results are evaluated by three key metrics: Hit Ratio at 500 (H@500), Normalized Discounted Cumulative Gain at 500 (NDCG@500), and Area Under the Curve (AUC), where higher values of all metrics indicate better model performance. As shown in Table \ref{tab:ablation_study}, the baseline model HyTRec, which lacks both the TADN branch (a linear attention with mixed attention structure) and the short-term attention branch (for learning short-term interest drift), achieves the lowest performance with H@500 of 0.6043, NDCG@500 of 0.3130, and AUC of 0.8355.

When only the short-term attention branch is introduced (while TADN is removed), the model performance is significantly improved, with H@500, NDCG@500, and AUC increasing to 0.6343, 0.3300, and 0.8505 respectively, demonstrating that capturing short-term interest drift can effectively enhance the model’s ability to perceive immediate user preferences. When only the TADN branch is retained (without short-term attention), the model achieves further performance improvement (H@500=0.6493, NDCG@500=0.3380, AUC=0.8575), indicating that the TADN branch with linear attention can efficiently model long-term behavioral sequence dependencies and outperforms the single short-term branch in overall recommendation accuracy. Notably, the full model integrating both TADN and short-term attention branches achieves the best performance across all metrics, with H@500 reaching 0.6643, NDCG@500 0.3480, and AUC 0.8655, which fully proves that the two branches complement each other, \textit{i.e.}, short-term attention captures immediate interest changes while TADN models long-term sequence patterns, jointly promoting the model’s performance on long behavioral sequence recommendation tasks.

\subsection{Efficiency Evaluation (RQ4)}
\begin{figure*}[t]
  \centering
  \includegraphics[width=0.99\textwidth]{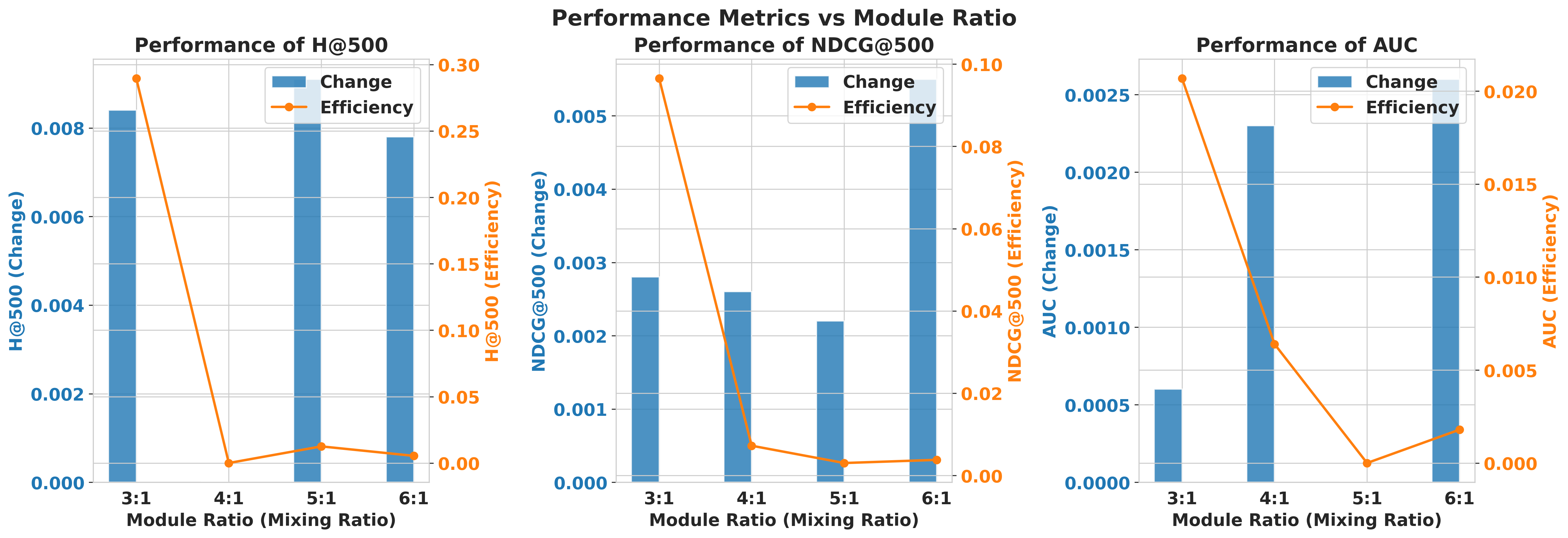}
  \caption{Performance Comparison Under Different Hybrid Attention Ratios.}
  \label{fig:efficiency_evaluation}
\end{figure*}

To evaluate the optimal ratio of the mixed attention model, we conducted comparative experiments with ratios ranging from 2:1 to 6:1. We calculated H@500, NDCG@500, AUC, and latency under different ratios, using the 2:1 ratio as the performance baseline. Efficiency is defined as the ratio of the performance change to the latency change, as shown in Figure \ref{fig:efficiency_evaluation}. 

As illustrated by the experimental results, the ratio of 3:1 yields relatively high efficiency gains across all three metrics, showing favorable overall performance. When the ratio is adjusted to 4:1, the efficiency gain on the first metric drops to zero, while the values on the other two metrics remain positive but decrease. For the ratio of 5:1, the efficiency value on the third metric becomes zero, with the other two metrics maintaining small positive improvements. In contrast, the ratio of 6:1 achieves positive efficiency values on all three metrics, but the overall gains are lower than those under the 3:1 setting. These observations demonstrate that an appropriate trade-off between linear attention and short-term attention is crucial for achieving optimal efficiency, and the 3:1 ratio provides the best balance between recommendation performance and inference efficiency.

\subsection{Case Study (RQ5)}
\begin{table}[t]
\centering
\caption{Model performance on several typical business bad cases.}
\label{tab:performance_badcases}
\setlength{\tabcolsep}{3.5pt} 
\resizebox{\linewidth}{!}{\begin{tabular}{lccc}
\toprule
\textbf{Badcase Type} & \textbf{H@500} & \textbf{NDCG@500} & \textbf{AUC} \\
\midrule
Cold-start for new users & 0.6622 & 0.3473 & 0.8616 \\
Silent old users          & 0.6340 & 0.2808 & 0.8661 \\
\bottomrule
\end{tabular}}
\end{table}
For users with sparse historical interactions, we categorize them into two types of challenging scenarios according to the distribution of their interaction information: new-user cold start and silent old users. We augment the historical behavior sequences of these two types of users by leveraging information from users with similar interests and relatively rich behavioral histories. Experimental results in Table \ref{tab:performance_badcases} demonstrate that HyTRec achieves superior performance in handling these two challenging scenarios, which can mainly be attributed to the similar decision-making patterns of similar user groups and the strong generalization ability of the proposed model.

\section{Conclusion}
This paper investigates how to endow behavior sequence models with the ability to efficiently process extremely long sequences. Our method first proposes an interactive item prediction model that integrates historical interests and short-term intentions, aiming to address the problem of user short-term interest drift that may exist in recommender systems. Specifically, we design a hybrid attention architecture for efficiently modeling long-term interests. Furthermore, we introduce a temporal decay mechanism and a short-term attention branch into the hybrid attention, thereby taking both users’ short-term intentions into account. Extensive experimental results strongly demonstrate the effectiveness of the proposed method with better efficiency.

\section*{Future Analysis}
To clarify the advantages of our proposed method (HyTRec) over existing models in cross-domain transfer tasks, Table \ref{tab:cross-domain} shows that our method outperforms the existing model SASRec significantly on all three key metrics, fully demonstrating our method’s remarkable ability to adapt to cross-domain data distribution differences and capture effective feature information—an advantage attributed to its rational structure that better overcomes domain shift impacts; based on these advantages and the existing research foundation, future work will focus on further enhancing the model’s cross-domain and cross-scenario generalization ability by optimizing the long-term encoder and temporal-aware delta network to reduce dependence on long-range context information and designing a more flexible adaptive mechanism, improving robustness against noisy sequential data through introducing noise detection/denoising mechanisms and a noise-aware sequential learning module, deepening research on the performance-efficiency trade-off in generative recommendation to balance high performance and efficiency via lightweight structures and efficient training strategies, and expanding the model’s application scope to more diverse cross-domain datasets and industrial scenarios to provide reliable technical support for practical cross-domain recommendation tasks.

\section*{Impact Statement}
We focus on designing efficient algorithms for generative recommendation. The datasets included in this paper are from publicly available materials with no sensitive information presented. Moreover, our efforts are solely devoted to research purposes with no intention for commercial use.
\bibliography{custom}

@article{de2015exploration,
  title={An exploration of softmax alternatives belonging to the spherical loss family},
  author={De Brebisson, Alexandre and Vincent, Pascal},
  journal={arXiv preprint arXiv:1511.05042},
  year={2015}
}

@inproceedings{qin2022devil,
  title={The devil in linear transformer},
  author={Qin, Zhen and Han, Xiaodong and Sun, Weixuan and Li, Dongxu and Kong, Lingpeng and Barnes, Nick and Zhong, Yiran},
  booktitle={Proceedings of the 2022 Conference on Empirical Methods in Natural Language Processing},
  pages={7025--7041},
  year={2022}
}

@article{yang2023qwen2,
  title={Qwen2 technical report},
  author={Yang, An and Yang, Baosong and Hui, Binyuan and Zheng, Bo and Yu, Bowen and Zhou, Chang and Li, Chengpeng and Li, Chengyuan and Liu, Dayiheng and Huang, Fei and others},
  journal={arXiv preprint arXiv:2407.10671},
  year={2024}
}

@inproceedings{pei2022transformer,
  title={Transformer uncertainty estimation with hierarchical stochastic attention},
  author={Pei, Jiahuan and Wang, Cheng and Szarvas, Gy{\"o}rgy},
  booktitle={Proceedings of the AAAI Conference on Artificial Intelligence},
  volume={36},
  number={10},
  pages={11147--11155},
  year={2022}
}

@inproceedings{hua2022transformer,
  title={Transformer quality in linear time},
  author={Hua, Weizhe and Dai, Zihang and Liu, Hanxiao and Le, Quoc},
  booktitle={International conference on machine learning},
  pages={9099--9117},
  year={2022},
  organization={PMLR}
}

@inproceedings{zhai2024actions,
  title={Actions Speak Louder than Words: Trillion-Parameter Sequential Transducers for Generative Recommendations},
  author={Zhai, Jiaqi and others},
  booktitle={International Conference on Machine Learning (ICML)},
  year={2024},
  note={Key baseline: HSTU}
}

@article{qin2025bridging,
  title={Bridging the Divide: Reconsidering Softmax and Linear Attention},
  author={Qin, Zhen and others},
  journal={arXiv preprint arXiv:2412.06590},
  year={2024},
  note={Accepted at ICLR 2025. Theory on Semantic Confusion/Injectivity.}
}

@inproceedings{ren2025long,
  title={Long-Sequence Recommendation Models Need Decoupled Embeddings},
  author={Ren, Rui and others},
  booktitle={International Conference on Learning Representations (ICLR)},
  year={2025}
}

@inproceedings{hidasi2016session,
  title={Session-based Recommendations with Recurrent Neural Networks},
  author={Hidasi, Bal{\'a}zs and Karatzoglou, Alexandros and Baltrunas, Linas and Tikk, Domonkos},
  booktitle={International Conference on Learning Representations (ICLR)},
  year={2016}
}

@inproceedings{kang2018self,
  title={Self-Attentive Sequential Recommendation},
  author={Kang, Wang-Cheng and McAuley, Julian},
  booktitle={IEEE International Conference on Data Mining (ICDM)},
  year={2018},
  pages={197--206}
}

@inproceedings{sun2019bert,
  title={BERT4Rec: Sequential Recommendation with Bidirectional Encoder Representations from Transformer},
  author={Sun, Fei and Liu, Jun and Wu, Jian and Pei, Changhua and Lin, Xiao and Ou, Wenwu and Jiang, Peng},
  booktitle={ACM International Conference on Information and Knowledge Management (CIKM)},
  year={2019},
  pages={1441--1450}
}

@inproceedings{geng2022recommendation,
  title={Recommendation as Language Processing (RLP): A Unified Pretrain, Personalized Prompt \& Predict Paradigm (P5)},
  author={Geng, Shijie and Liu, Shuchang and Fu, Zuohui and Ge, Yingqiang and Zhang, Yongfeng},
  booktitle={ACM Recommender Systems (RecSys)},
  year={2022},
  pages={299--315}
}

@inproceedings{bao2023tallrec,
  title={TALLRec: An Effective and Efficient Tuning Framework to Align Large Language Model with Recommendation},
  author={Bao, Keqin and Zhang, Jizhi and Zhang, Yang and Wang, Wenjie and Feng, Fuli and He, Xiangnan},
  booktitle={ACM International Conference on Recommender Systems (RecSys)},
  year={2023}
}

@article{lin2025survey,
  title={A Survey on LLM-powered Agents for Recommender Systems},
  author={Lin, Jiayu and others},
  journal={arXiv preprint arXiv:2311.13375},
  year={2025}
}

@inproceedings{vaswani2017attention,
  title={Attention is all you need},
  author={Vaswani, Ashish and others},
  booktitle={Advances in Neural Information Processing Systems (NIPS)},
  year={2017}
}

@article{beltagy2020longformer,
  title={Longformer: The long-document transformer},
  author={Beltagy, Iz and Peters, Matthew E and Cohan, Arman},
  journal={arXiv preprint arXiv:2004.05150},
  year={2020}
}

@inproceedings{choromanski2020rethinking,
  title={Rethinking Attention with Performers},
  author={Choromanski, Krzysztof and others},
  booktitle={International Conference on Learning Representations (ICLR)},
  year={2021}
}

@inproceedings{gu2021efficiently,
  title={Efficiently modeling long sequences with structured state spaces},
  author={Gu, Albert and Goel, Karan and Re, Christopher},
  booktitle={International Conference on Learning Representations (ICLR)},
  year={2022},
  note={S4 Model}
}

@article{gu2023mamba,
  title={Mamba: Linear-Time Sequence Modeling with Selective State Spaces},
  author={Gu, Albert and Dao, Tri},
  journal={arXiv preprint arXiv:2312.00752},
  year={2023}
}

@inproceedings{yang2024parallelizing,
  title={Parallelizing Linear Transformers with the Delta Rule over Sequence Length},
  author={Yang, Songlin and Wang, Bailin and Zhang, Yu and Shen, Yikang and Kim, Yoon},
  booktitle={Advances in Neural Information Processing Systems (NeurIPS)},
  year={2024},
  note={DeltaNet / Linear V2}
}

@inproceedings{yang2024gated,
  title={Gated Linear Attention Transformers with Hardware-Efficient Training},
  author={Yang, Songlin and Wang, Bailin and Shen, Yikang and Panda, Rameswar and Kim, Yoon},
  booktitle={International Conference on Machine Learning (ICML)},
  year={2024},
  note={Gated Linear Attention (GLA)}
}

@article{lieber2024jamba,
  title={Jamba: A Hybrid Transformer-Mamba Language Model},
  author={Lieber, Opher and others},
  journal={arXiv preprint arXiv:2403.19887},
  year={2024},
  note={Key paper for Hybrid Architecture trend}
}

@inproceedings{pi2020search,
  title={Search-based User Interest Modeling with Long-term Sequential Behavior Data for Click-Through Rate Prediction},
  author={Pi, Qi and others},
  booktitle={Proceedings of the 26th ACM SIGKDD International Conference on Knowledge Discovery \& Data Mining},
  pages={2494--2503},
  year={2020},
  note={SIM model from Alibaba}
}

@inproceedings{chen2021end,
  title={End-to-End User Behavior Retrieval in Click-Through Rate Prediction Model},
  author={Chen, Jiarui and others},
  booktitle={Proceedings of the 44th International ACM SIGIR Conference on Research and Development in Information Retrieval},
  pages={1207--1216},
  year={2021},
  note={ETA model using LSH}
}

@inproceedings{li2019multi,
  title={Multi-interest Network with Dynamic Routing for Recommendation at Tmall},
  author={Li, Chao and others},
  booktitle={Proceedings of the 28th ACM International Conference on Information and Knowledge Management},
  pages={2615--2623},
  year={2019},
  note={MIND model}
}

@inproceedings{cen2020controllable,
  title={Controllable Multi-interest Framework for Recommendation},
  author={Cen, Yukuo and others},
  booktitle={Proceedings of the 26th ACM SIGKDD International Conference on Knowledge Discovery \& Data Mining},
  pages={2946--2956},
  year={2020},
  note={ComiRec}
}

@inproceedings{press2021train,
  title={Train Short, Test Long: Attention with Linear Biases Enables Input Length Extrapolation},
  author={Press, Ofir and Smith, Noah A and Lewis, Mike},
  booktitle={International Conference on Learning Representations (ICLR)},
  year={2022},
  note={ALiBi - Basis for decay gating}
}

@article{wang2025simuser,
  title={SimUSER: Simulating User Behavior with Large Language Models for Recommender System Evaluation},
  author={Wang, Xinyi and others},
  journal={arXiv preprint arXiv:2504.12722},
  year={2025}
}

@article{liu2025tablepilot,
  title={TablePilot: Recommending Human-Preferred Tabular Data Analysis with Large Language Models},
  author={Liu, Haotian and others},
  journal={arXiv preprint arXiv:2503.13262},
  year={2025}
}

@inproceedings{shao2025motir,
  title={MotiR: Motivation-aware Retrieval for Long-Tail Recommendation},
  author={Shao, Zeren and others},
  booktitle={Proceedings of the 63rd Annual Meeting of the Association for Computational Linguistics (ACL Industry Track)},
  year={2025}
}

@inproceedings{shukla2025reinforcement,
  title={Reinforcement Learning for Adversarial Query Generation to Enhance Relevance in Cold-Start Product Search},
  author={Shukla, Rishabh and others},
  booktitle={Proceedings of the 63rd Annual Meeting of the Association for Computational Linguistics (ACL Industry Track)},
  year={2025}
}

@inproceedings{zhou2018deep,
  title={Deep Interest Network for Click-Through Rate Prediction},
  author={Zhou, Guorui and Zhu, Xiaoqiang and Song, Chengru and Fan, Ying and Zhu, Han and Ma, Xiao and Yan, Yanghui and Jin, Junqi and Li, Han and Gai, Kun},
  booktitle={Proceedings of the 24th ACM SIGKDD International Conference on Knowledge Discovery \& Data Mining},
  pages={1059--1068},
  year={2018}
}

@article{deng2025onerec,
  title={OneRec: Unifying retrieve and rank with generative recommender and iterative preference alignment},
  author={Deng, Jin and Wang, S and Cai, K and Ren, L and Hu, Q and Ding, W and Luo, Q and Zhou, G},
  journal={arXiv preprint arXiv:2502.18965},
  year={2025}
}

@article{du2025mom,
  title={MoM: Linear sequence modeling with mixture-of-memories},
  author={Du, J and Sun, W and Lan, D and Hu, J and Cheng, Y},
  journal={arXiv preprint arXiv:2502.13685},
  year={2025}
}

@inproceedings{gu2024mamba,
  title={Mamba: Linear-time sequence modeling with selective state spaces},
  author={Gu, Albert and Dao, Tri},
  booktitle={First Conference on Language Modeling},
  year={2024}
}

@article{sun2024learning,
  title={Learning to (learn at test time): RNNs with expressive hidden states},
  author={Sun, Y and Li, X and Dalal, K and Xu, J and Vikram, A and Zhang, G and Dubois, Y and Chen, X and Wang, X and Koyejo, S and others},
  journal={arXiv preprint arXiv:2407.04620},
  year={2024}
}

@inproceedings{fakehr1,
      title={Fake-HR1: Rethinking Reasoning of Vision Language Model for Synthetic Image Detection}, 
      author={Changjiang Jiang and Xinkuan Sha and Fengchang Yu and Jingjing Liu and Jian Liu and Mingqi Fang and Chenfeng Zhang and Wei Lu},
      year={2026},
      booktitle={ICASSP},
      url={https://arxiv.org/abs/2602.10042} 
}

@inproceedings{tabdsr,
  title={TABDSR: Decompose, Sanitize, and Reason for Complex Numerical Reasoning in Tabular Data},
  author={Jiang, Changjiang and Yu, Fengchang and Chen, Haihua and Lu, Wei and Zeng, Jin},
  booktitle={Findings of the Association for Computational Linguistics: EMNLP 2025},
  pages={3172--3196},
  year={2025}
}

\appendix

\label{sec:appendix}
\twocolumn
\section{The Evolution of Attention Mechanisms}

\subsection{Softmax Attention}
Softmax attention (scaled dot-product attention) is the core operation in Transformers
\cite{vaswani2017attention}. It computes pairwise similarities between queries and keys, applies a
softmax normalization, and aggregates values:
\begin{equation}
O = \text{Softmax}\left( \frac{QK^\top}{\sqrt{d}} \right)V,
\end{equation}
where $Q,K,V\in\mathbb{R}^{n\times d}$, $n$ is the sequence length and $d$ is the hidden
dimension. Its time and memory complexity are quadratic in $n$ (e.g., $O(n^2d)$ time and storing an
$n\times n$ attention matrix), which limits scalability for long sequences.

\subsection{Linear Attention}
Linear attention aims to reduce the quadratic cost of softmax attention by avoiding explicit
materialization of the $n\times n$ attention matrix, typically via kernelization and reordering
matrix multiplications \cite{de2015exploration}. A representative form can be written as
\begin{equation}
O = \text{Norm}\left( Q(K^\top V) \right),
\end{equation}
which yields complexity linear in sequence length (e.g., $O(nd^2)$). For causal settings, linear
attention can be implemented with recurrent state updates, but maintaining causality may introduce
additional scan/cumsum-style operations that hurt parallel efficiency \cite{hua2022transformer}.
We follow the common observation that purely linear variants may sacrifice retrieval fidelity, as
also discussed in recent long-context architectures such as TransNormer \cite{qin2022devil}.

\subsection{Hybrid Attention}
Hybrid attention combines the efficiency of linear attention with the retrieval quality of
softmax attention by inserting softmax-attention layers sparsely among linear layers. For a model
with $l$ layers, let $L=\{1,2,\dots,l\}$ be the layer indices and $\mathcal{S}\subseteq L$ denote
the layers using softmax attention. For each layer $i\in L$:

\begin{equation}
O_i = \begin{cases}
\text{SoftmaxAttention} \\ 
\quad (Q_i, K_i, V_i), & i \in \mathcal{S} \\[1.5ex]
\text{LinearAttention} \\ 
\quad (Q_i, K_i, V_i), & \text{otherwise}
\end{cases}
\end{equation}

With a sparse $\mathcal{S}$, the overall complexity remains near-linear in $n$ while improving
retrieval compared to purely linear backbones, which is the motivation behind HyTRec.

\begin{figure*}[t]
  \centering
  \includegraphics[width=0.99\textwidth]{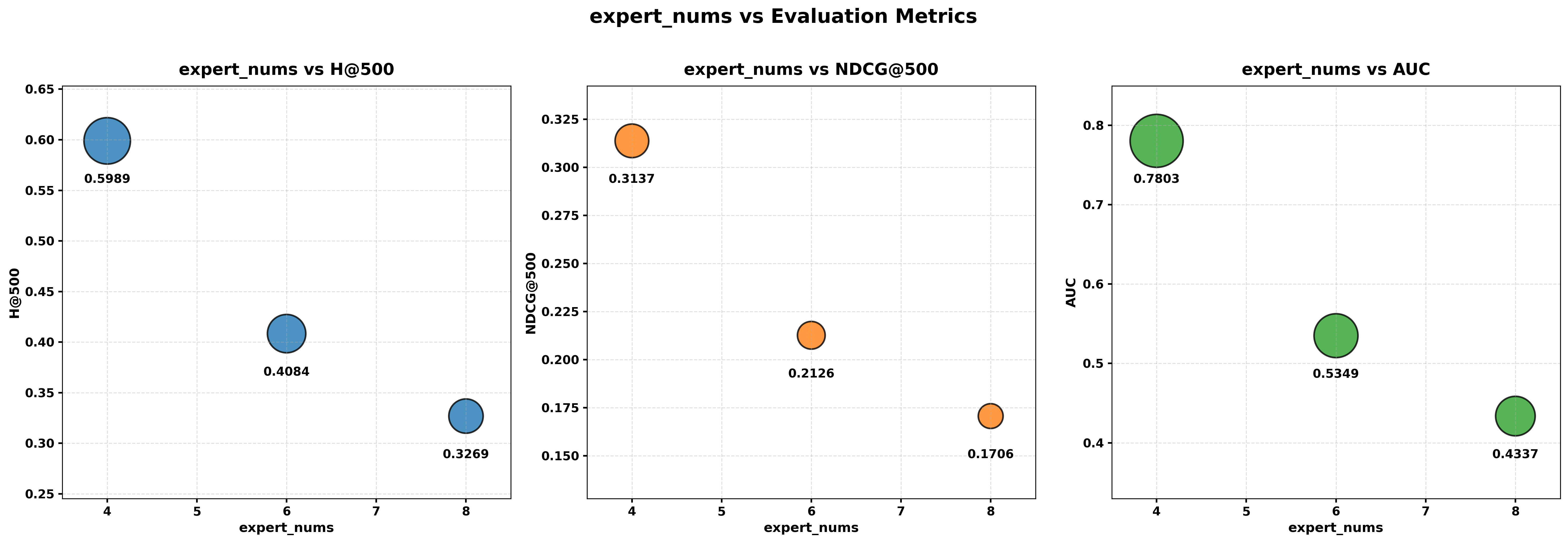}
  \caption{Performance Comparison Under Different Number of Experts}
  \label{fig:expert_nums_evaluation}
\end{figure*}

\section{Baseline Model Descriptions}
\subsection{Behavior Sequence Models}
These models primarily focus on modeling user behavior sequences to capture sequential preferences for recommendation tasks.

\noindent \textbf{GRU4Rec} \cite{hidasi2016session}: A canonical sequential recommendation model based on Gated Recurrent Units (GRUs). It models user behavior sequences as time-series data and uses recurrent neural networks to capture temporal dependencies.

\begin{itemize}
  \item Core Architecture: GRU-based encoder for sequential behavior encoding.
  \item Key Mechanism: Utilizes GRU's gating mechanism to control the flow of historical information, capturing short-term and long-term user preferences.
\end{itemize}

\noindent \textbf{SASRec} \cite{kang2018self}: Introduces self-attention mechanism into sequential recommendation, replacing recurrent structures with transformer-style self-attention to model long-range dependencies in behavior sequences.

\begin{itemize}
  \item Core Architecture: Transformer encoder (multi-head self-attention with feed-forward network).
  \item Key Mechanism: Causal self-attention mask to ensure only historical behaviors are used for prediction, avoiding information leakage.
\end{itemize}

\noindent \textbf{DIN} \cite{zhou2018deep}:  Deep Interest Network (DIN) focuses on capturing dynamic user interests by designing an attention-based interest activation mechanism, adapting to different candidate items.

\begin{itemize}
  \item Core Architecture: Embedding layer, interest extraction network, and attention-based interest activation.
  \item Key Mechanism: Item-aware attention weights to dynamically aggregate user historical interests based on the target item.
\end{itemize}

\noindent \textbf{HSTU} \cite{pei2022transformer}: Hierarchical Sequential Transformer with Uncertainty (HSTU) enhances sequential recommendation by incorporating hierarchical structure and uncertainty modeling into transformer-based sequential models.

\begin{itemize}
  \item Core Architecture: Hierarchical transformer encoder (behavior-level with session-level).
  \item Key Mechanism: Uncertainty estimation module to quantify the reliability of sequential patterns, improving robustness.
\end{itemize}

\begin{figure*}[t]
  \centering
  \includegraphics[width=0.99\textwidth]{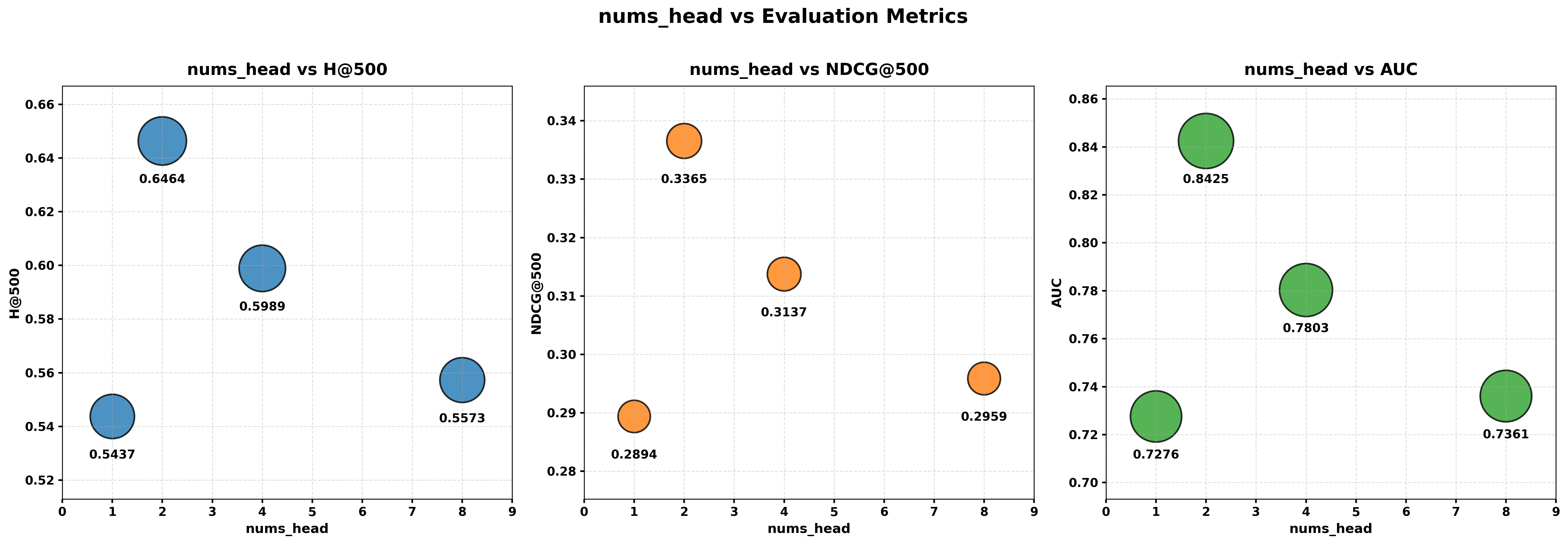}
  \caption{Performance Comparison Under Different Number of Attention Heads}
  \label{fig:nums_head_evaluation}
\end{figure*}

\subsection{Long-Text Models}
These models are designed to handle long-text inputs, with optimized attention mechanisms or model structures to address the computational bottleneck of vanilla transformers on long sequences.

\noindent \textbf{Transformer} \cite{vaswani2017attention}: The foundational model with scaled dot-product attention, serving as the baseline for long-text modeling despite its quadratic computational complexity.

\begin{itemize}
  \item Core Architecture: Multi-head self-attention with position-wise feed-forward network.
  \item Key Mechanism: Scaled dot-product attention (softmax attention) to model global dependencies, with positional encoding for sequence order.
\end{itemize}

\noindent \textbf{GLA} \cite{yang2024gated}: Gated Linear Attention (GLA) is a variant of linear attention with data-dependent gating mechanisms, reducing the computational complexity from quadratic to linear in sequence length.

\begin{itemize}
  \item Core Architecture: linear attention framework with gating modules.
  \item Key Mechanism: Diagonal decay matrix ($A_t = \text{diag}(\alpha_t)$) for forgetting and input-dependent gate ($\beta_t$) for injection, enabling efficient long-sequence processing.
\end{itemize}

\noindent \textbf{Qwen-next} \cite{yang2023qwen2}: Qwen-next is an optimized version of the Qwen large language model tailored for long-text understanding, using 2-block structured design to balance efficiency and performance.

\begin{itemize}
  \item Core Architecture: Modified transformer with block-wise attention.
  \item Key Mechanism: 2-block partitioning strategy for long-text encoding, reducing I/O overhead and computational cost while maintaining contextual modeling ability.
\end{itemize}

\subsection{Contrast with HyTRec}
HyTRec occupies a unique middle ground. It inherits the high accuracy of softmax attention while maintaining the efficiency of linear attention, and incorporates the hybrid attention structure with prior knowledge from the recommendation domain, as shown in Table \ref{tab:model_comparison}.

\begin{table*}[t]
  \centering
  \setlength{\tabcolsep}{5pt}
  \caption{Comparison of different models on key evaluation metrics. HyTRec is the first work in the recommendation field to achieve a balance between efficiency and accuracy based on hybrid attention, enabling effective modeling of long user behavior sequences.}
  \resizebox{0.8\textwidth}{!}{
    \begin{tabular}{lcccc}
      \toprule
      \textbf{Model} & \textbf{Efficiency} & \textbf{Temporal Info} & \textbf{Short-term Intent} & \textbf{Rec Capability} \\
      \midrule
      Softmax Attention & $\times$ & $\times$ & $\times$ & $\times$ \\ 
      Linear Attention  & $\checkmark$ & $\times$ & $\times$ & $\times$ \\
      Qwen-next         & $\checkmark$ & $\times$ & $\times$ & $\times$ \\
      \textbf{Ours}     & $\checkmark$ & $\checkmark$ & $\checkmark$ & $\checkmark$ \\
      \bottomrule
    \end{tabular}
  }
  \vspace{-5pt}
  \label{tab:model_comparison}
\end{table*}
\section{Engineering Tricks for Long Behavior Sequence Construction}
\subsection{Extending the Time Window Size}

For data associated with the same user ID, merge data from different partitions in the backend data warehouse. This operation allows tracing user behaviors back to the time of their registration, significantly improving the utilization of long-sequence data.

\subsection{Intermediate Process Sequence Data Processing Strategy}
Treat the ad attribution ID as a numerical identifier to trace the user's states throughout the intermediate funnel (ad click, redirect, activation, wake-up, product click, stay, redirect, collection, add-to-cart, payment). Sort different behavior data based on their correlation strength with payment behavior to form hierarchical long-cycle sequence data.
\begin{table*}[t]
  \centering
  \caption{Comparison of different nums\_head on key recommendation evaluation metrics. The table shows the performance (H@500, NDCG@500, AUC) and efficiency (Latency) under different nums\_head settings.}
  \begin{tabular}{lcccc}
    \toprule
    \textbf{nums\_head} & \textbf{H@500} & \textbf{NDCG@500} & \textbf{AUC} & \textbf{Latency} \\
    \midrule
    1 & 0.6514 & 0.3467 & 0.8717 & 1.1981 \\
    2 & 0.6623 & 0.3459 & 0.8634 & 1.1308 \\
    \textbf{4} & \textbf{0.6643} & \textbf{0.3480} & \textbf{0.8655} & \textbf{1.1092} \\
    8 & 0.6579 & 0.3493 & 0.8690 & 1.1806 \\
    \bottomrule
  \end{tabular}
  \label{tab:nums_head_comparison}
\end{table*}
\begin{table*}[htbp]
  \centering
  \caption{Comparison of different expert\_nums on key recommendation evaluation metrics. The table presents the performance and latency under various expert number settings.}
  \begin{tabular}{lcccc}
    \toprule
    \textbf{Expert\_nums} & \textbf{H@500} & \textbf{NDCG@500} & \textbf{AUC} & \textbf{Latency} \\
    \midrule
    \textbf{4} & \textbf{0.6643} & \textbf{0.3480} & \textbf{0.8655} & \textbf{1.1092} \\
    6 & 0.6624 & 0.3449 & 0.8677 & 1.6221 \\
    8 & 0.6579 & 0.3434 & 0.8729 & 2.0128 \\
    \bottomrule
  \end{tabular}
  \label{tab:expert_nums_comparison}
\end{table*}
\begin{table*}[htbp]
  \centering
  \caption{Comparison of different Ratio on key recommendation evaluation metrics. The table presents the performance and latency under various ratio settings.}
  \begin{tabular}{lcccc}
    \toprule
    \textbf{Ratio} & \textbf{H@500} & \textbf{NDCG@500} & \textbf{AUC} & \textbf{Latency} \\
    \midrule
    2:1 & 0.6559 & 0.3452 & 0.8649 & 1.0802 \\
    \textbf{3:1} & \textbf{0.6643} & \textbf{0.3480} & \textbf{0.8655} & \textbf{1.1092} \\
    4:1 & 0.6527 & 0.3478 & 0.8672 & 1.4397 \\
    5:1 & 0.6650 & 0.3474 & 0.8619 & 1.8070 \\
    6:1 & 0.6637 & 0.3507 & 0.8675 & 2.5296 \\
    \bottomrule
  \end{tabular}
  \label{tab:ratio_comparison}
\end{table*}
\subsection{Rational Utilization of Different Channels}
For a single e-commerce platform, only product data can be utilized. For community-ecommerce platforms, community data can be used to enhance product behavior data. Specifically, information extraction technology can be employed to identify users' interests in different product categories from community articles.

\subsection{Missing Value Imputation}

In real-world e-commerce business data, short-sequence samples account for a high proportion, while long-sequence samples are usually classified as active users or even high-value active users. As high-stickiness users, they are often regarded as ``easy samples'' with low modeling difficulty.
An effective modeling approach is to impute the missing historical behaviors and empty time intervals without interactions in short-sequence samples based on user stratification strategies. This can greatly improve the effective information of ``hard samples'', thereby better enhancing model performance and gains in business metrics.

\section{Common Engineering Bad Cases}

In some typical business bad cases, the scenarios themselves are not suitable for direct long-sequence modeling. Appropriate data strategies must be adopted to enable such samples to provide positive feedback to the overall performance of long-sequence modeling.

\noindent \textbf{(1) Cold start for new users:} Such samples have extremely sparse interactions or even zero interactions, and basically have no valid behavior sequences available for modeling.

\noindent \textbf{(2) Inactive long-term users:} Such samples only have a small number of interaction records from a very long time ago, or only complete registration without subsequent behaviors.

\noindent \textbf{(3) Scalper accounts:} Such samples are often misidentified as high-value users due to their massive interaction records, which seem to be ideal samples for long-sequence modeling. However, modeling this group of users hardly brings positive contributions to platform revenue.

\section{More Experimental Results}

\subsection{Comparison Experiments on Different Numbers of Attention Heads}
In the vanilla Transformer block, the number of heads is usually set to an even number. To achieve efficient behavior sequence modeling, we evaluate the model performance under different numbers of attention heads: 2, 4, 6, and 8.

As shown in Table \ref{tab:nums_head_comparison} and Figure \ref{fig:nums_head_evaluation}, when the number of attention heads increases from 1 to 8, the model performance first improves slightly and then declines gradually, while the inference latency exhibits a trend of decreasing first and then increasing. Considering both the recommendation performance metrics and inference efficiency, setting the number of attention heads to 2 is the overall optimal parameter choice.

\subsection{Comparison Experiments on Different Numbers of Experts}

In recommendation systems based on the multi-expert structure, the number of experts is generally related to the heterogeneity of user groups. According to business prior knowledge, users can be divided into four heterogeneous groups: new users, old users, silent old users, and churned users. Therefore, we set the number of experts to 4, 6, and 8 for comparison experiments, corresponding to the number of user groups, 1.5 times, and 2 times the number of groups respectively, so as to verify the impact of the number of experts on model performance.

As shown in Table \ref{tab:expert_nums_comparison} and Figure \ref{fig:expert_nums_evaluation}, when the number of experts increases from 4 to 8, the recommendation performance of the model continues to decline, while the inference latency rises significantly. Considering the recommendation metrics H@500, NDCG@500, AUC and inference latency, setting the number of experts to 4 achieves the overall optimal performance.

\subsection{Empirical Analysis of Hybrid Design}

Table \ref{tab:ratio_comparison} illustrates the impact of different hybrid structures (Ratio) on model performance and inference latency. It can be observed that the 2:1 structure achieves the lowest latency (1.0802), but there is still room for improvement in various evaluation metrics. When the ratio is increased to 3:1, the model achieves high scores in H@500, NDCG@500, and AUC with no significant extra latency overhead, reaching the optimal balance between accuracy and efficiency. As the ratio continues to increase to 4:1 and 5:1, the latency rises noticeably despite slight fluctuations or minor improvements in some metrics. When the ratio reaches 6:1, although NDCG@500 and AUC achieve their peak values, the latency surges to 2.5296, resulting in excessively high computational cost. In summary, 3:1 is the optimal trade-off configuration between performance and efficiency. It can stably improve the core recommendation metrics including H@500, NDCG@500, and AUC while controlling latency cost, which fully validates the effectiveness and practicality of the proposed hybrid structure design.

\section{Discussion}
In this work, we propose HyTRec, a hybrid attention framework for generative recommendation, which integrates linear attention and softmax attention to address the core challenge of long sequence modeling in recommendation systems—balancing efficiency and semantic integrity. While HyTRec effectively mitigates the efficiency-semantic dilemma faced by traditional attention-based recommendation models, it still has room for improvement in further enhancing performance and adaptability. Here, we first briefly recap the structural trade-offs and core advantages of the proposed hybrid attention mechanism, and then focus on outlining four key potential directions for future research to address the existing limitations of HyTRec, all in line with the analytical logic of linear sequence modeling discussions.

\subsection{Adaptive Boundary for Hybrid Attention}
One key limitation of HyTRec’s current hybrid attention design lies in the empirical setting of the boundary between long historical interactions and recent interactions, which lacks adaptability to diverse user groups. To address this limitation, future work should focus on designing an adaptive boundary adjustment mechanism for the hybrid attention module. Instead of using a fixed threshold to separate long historical and recent interactions, we can leverage user-specific characteristics (\textit{e.g.}, interaction frequency, intent stability) to dynamically allocate the proportion of linear attention and softmax attention. For example, for users with stable long-term preferences, we can increase the scope of linear attention to reduce computational cost; for users with frequent intent shifts, we can expand the scope of softmax attention to capture fine-grained intent changes.

\subsection{Integration with Expanded Memory Architectures}

Similar to the memory capacity wall challenge faced by linear recurrence models, HyTRec’s linear attention module still suffers from memory overwriting in fixed-dimensional states when processing extremely long interaction sequences (\textit{e.g.}, 10k+ tokens), which limits its performance in handling users with extensive interaction histories. To break through this constraint, future research can integrate HyTRec with expanded memory architectures to explicitly expand the model’s memory capacity. This integration will allow the hybrid attention module to retain more valuable historical information while maintaining linear efficiency, further improving performance for users with extensive interaction histories.

\subsection{Extension to Multi-Scenario Recommendation}

Currently, HyTRec is only validated on e-commerce recommendation datasets, which restricts its generalizability to other generative recommendation scenarios. To expand the applicability of HyTRec, future work can extend the framework to other generative recommendation scenarios, such as content recommendation (\textit{e.g.}, articles, videos) and social recommendation. In these scenarios, the characteristics of user interactions (\textit{e.g.}, interaction frequency, semantic complexity) differ significantly from e-commerce, which will require adjusting the hybrid attention design and TADN to adapt to scenario-specific needs. Additionally, verifying the model’s performance on cross-domain recommendation tasks can further demonstrate its generalizability.
\end{document}